\documentclass{article}
\usepackage{caption}
\usepackage{hyperref}
\usepackage{float} 
\pdfoutput=1
\usepackage{subfig}
\usepackage{array}
\usepackage{cite}
\usepackage{amsmath,amssymb,amsfonts}
\usepackage{algorithmic}
\usepackage{graphicx}
\usepackage{textcomp}
\usepackage{xcolor}
\usepackage{makecell}
\usepackage{tikz}
\usepackage{tikz}
\usepackage{multicol}
\usepackage{authblk} 
\usepackage{multicol}
\usetikzlibrary{decorations.pathreplacing}
\def\BibTeX{{\rm B\kern-.05em{\sc i\kern-.025em b}\kern-.08em
    T\kern-.1667em\lower.7ex\hbox{E}\kern-.125emX}}
\begin{document}

\title{Semantic Communication-Enabled Cloud-Edge-End-collaborative Metaverse Services Architecure\\
	
	\thanks{Digital Object Identifier 10.1109/TVT.2024.3427713}
}

\author[1]{Yuxuan Li}
\author[2]{Sheng Jiang}
\author[3]{Bizhu Wang}

\affil[ ]{Key Laboratory of Networking and Switching Technology}
\affil[ ]{Beijing University of Posts and Telecommunications, Beijing, China}
\affil[ ]{Emails: \{\texttt{liyuxuan@bupt.edu.cn}, \texttt{shengjiang@bupt.edu.cn}, \texttt{wangbizhu\_7@bupt.edu.cn}\}}
%\affil[ ]{Emails:{liyuxuan, shengjiang, wangbizhu\_7}@bupt.edu.cn}
\maketitle
\begin{abstract}
	
With technology advancing and the pursuit of new audiovisual experiences strengthening, the metaverse has gained surging enthusiasm. However, it faces practical hurdles as substantial data like high-resolution virtual scenes must be transmitted between cloud platforms and VR devices. Specifically, the VR device's wireless transmission hampered by insufficient bandwidth, causes speed and delay problems. Meanwhile, poor channel quality leads to data errors and worsens user experience. To solve this, we've proposed the Semantic Communication-Enabled Cloud-Edge-End Collaborative Immersive Metaverse Service (SC-CEE-Meta) Architecture, which includes three modules: VR video semantic transmission, video synthesis, and 3D virtual scene reconstruction. By deploying semantic modules on VR devices and edge servers and sending key semantic info instead of focusing on bit-level reconstruction, it can cut latency, resolve the resource-bandwidth conflict, and better withstand channel interference. Also, the cloud deploys video synthesis and 3D scene reconstruction preprocessing, while edge devices host 3D reconstruction rendering modules, all for immersive services. Verified on Meta Quest Pro, the SC-CEE-Meta can reduce wireless transmission delay by 96.05\% and boost image quality by 43.99\% under poor channel condition.

\end{abstract}

\section{Introduction}
With the continuous progress of communication network, artificial intelligence, graphics processing and other technologies, as well as people's increasing pursuit new audio-visual experience, metaverse services have once again set off a wave of enthusiasm \cite{cheng2023metaverse , alsamhi2024multisensory , damar2021metaverse}. Once users put on VR headsets, their devices will connect to the metaverse platform. Data is transmitted wirelessly (with wired connections also supported in some cases for more stable transmission), such as sending users' real-time videos, movements and
operation instructions to the server and receiving data like virtual scene images in return. This makes users feel immersed in a virtual 3D world, where they can freely explore, socialize and engage in various activities.\cite{yang2023survey,chang20226g}.

However, it faces practical obstacles as substantial data transmission between cloud platforms and VR devices burdens the network. When using wireless VR devices, there's a significant gap between available wireless bandwidth (5G's global average download bandwidth is only about 160 Mbps \cite{zhang2021key}) and the bandwidth needed for high-quality VR video (120 frames of 8K video requires at \cite{hazarika2023towards}),meaning the large video data for metaverse often exceeds wireless capacity. Moreover, the wireless environment is prone to interferences and uncertainties. Poor channel quality, influenced by signal obstructions, surrounding electromagnetic interference and signal fluctuations, causes data errors like packet loss or corruption, worsening the metaverse user experience and posing huge challenges for metaverse service deployment, as users may face lagging visuals, interrupted interactions and a lack of seamless immersion.

To tackle the challenges posed by wireless content transmission in the metaverse, several previous studies have centered on leveraging millimeter waves to support the Metaverse . Additionally, they have explored combining millimeter waves with alternative wireless systems like visible light communication or sub-6GHz bands as a complementary approach to boost the performance of mmWave technology . Nevertheless, despite these efforts, the inherent propagation characteristics of millimeter wave band signals render them highly vulnerable to being obstructed and attenuated by obstacles during the transmission process. As a consequence, this leads to limited signal coverage area and a significant increase in the uncertainty regarding transmission delay. What's more, these existing solutions have failed to fundamentally resolve the contradiction between the extensive data transmission requirements of the metaverse and the limited availability of wireless resources, thus still leaving room for improvement in ensuring smooth and efficient metaverse services.

Given the limitations of the existing metaverse service, we propose a semantic communication enabled cloud-edge-end colaborative service framework to provide high-quality immersive metaverse (SC-CEE-Meta) services even under circumstances of limited wireless resources and poor channel conditions. The main contributions are summarized as follows:

1. We proposed the SC-CEE-Meta framework, which integrates semantic
communication technology and a cloud-edge-end collaborative architecture to offer high-quality, immersive metaverse services despite limited wireless resources and poor channel conditions.

2. The SC-CEE-Meta Framework consists of three modules: VR video semantic
transmission (VSC), video synthesis (VS), and 3D virtual scene reconstruction (VSR). Deploying semantic modules on VR and edge servers and transmitting key semantic info instead of focusing on bit-level reconstruction helps cut latency, resolve the resource-bandwidth
conflict, and handle channel interference. The cloud handles video synthesis and 3D scene reconstruction preprocessing, while edge devices have rendering modules, all for immersive services.

3. Verified on the Quest Pro, the SC-CEE-Meta supports functions like video synthesis , 3D scene reconstruction, and rendering, and can reduce wireless transmission delay by 96.05\% and improve image quality by 43.99\% under poor channel quality.

\section{ System model}
The proposed SC-CEE-Meta framework simulates a user using a VR device to watch a 3D composite video of his or her virtual avatar in a distant background.
As shown in the figure \ref{fig:arc}, taking the user's request to watch a 3D composite video of his or her own video in a distant background as an example, the metaverse service process of the proposed framework is as follows:

\begin{figure}
	\centering
	\includegraphics[width=0.9\linewidth]{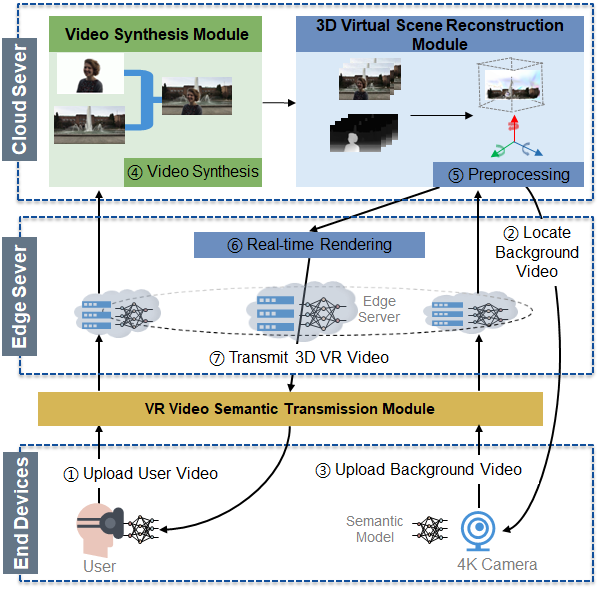}
	\caption{Diagram of the proposed SC-CEE-Met module, including the cloud layer, the edge layer, and the end layer.}
	\label{fig:arc}
\end{figure}

As shown in Fig. \ref{fig:arc}, when a user requests a service, the VR device uploads the user video via the VSC\footnote{with semantic encoding and decoding parts deployable on the sending and receiving ends} module to the edge server. The edge server restores the video using VSC module and sends it with the request to the cloud server.
Next, the cloud quickly locates the required distinct background video through the intelligent scheduling system. Similar to the user video, by applying the VSC module on the 4K camera and edge server, the background video is efficiently transmitted.
Then, the cloud invokes the VS module to generate a synthetic video from the user and background videos. The synthesis video is passed to the VSR module for 3D reconstruction to provide immersive service. The VSR module involves two steps: computationally intensive preprocessing in the cloud and real-time rendering on the edge server.
Finally, the 3D VR video is transmitted from the edge server to the VR device by invoking the VSC module, enabling users to enjoy the Metaverse service smoothly on VR devices.
\begin{table}[htbp]
	\captionsetup{font={tiny}, justification=raggedright}
	\caption{DENOTATIONS AND MEANINGS}
	\begin{center}
		\begin{tabular}{cc}
			\hline
			Denotation & Explanations \\
			\hline
			$i$ & video frame \\
			$B^{i}$ & background video \\
			$X^{i}$ & real-time video \\
			$\hat{X}^{i}$ & real-time video through semantic communication \\
			$\hat{B}^{i}$ & background video through semantic communication \\
			$I^{i}$ & fusion video \\
			$X_{g}$ & a GOP which contains certain frames \\
			$S$ & MODnet's low-resolution branch \\
			$D$ & MODNet's high-resolution branch \\
			$F$ & MODnet's fusion branch \\
			$d_{\alpha}$ & boundary detail matte \\
			$m_{d}$ & binary mask focus on the portrait boundaries \\
			$s_{p}$ & predicted coarse semantic mask \\
			$K_{t}$ & camera intrinsic \\
			$I_{t}$ & input frame \\
			$\mu _{0},R_{0}$ & 3D mean and orientation in the canonical frame \\
			$ $ & in the canonical frame \\
			$E,K$ & world-to-camera extrinsic and intrinsic \\
			$K$ & world-to-camera \\
			$\Pi $ & perspective projection \\
			$J_{KE} $ & Jacobian of perspective projection the point $\mu _{0}$ \\
			$p$ & pixel \\
			$o_{i}$ & opacity \\
			$(·)_{[3]}$ & the third element of a vector \\
			$\mathbb{S}\mathbb{E}(3)$ & motion bases \\
			$\hat{D}$ & depth value \\
			\hline
		\end{tabular}
		\label{tab1}
	\end{center}
\end{table}
\subsection{VR Video Semantic Transmission (VST) Module}
\begin{figure}[tbph]
	\centering
	\includegraphics[width=1\linewidth]{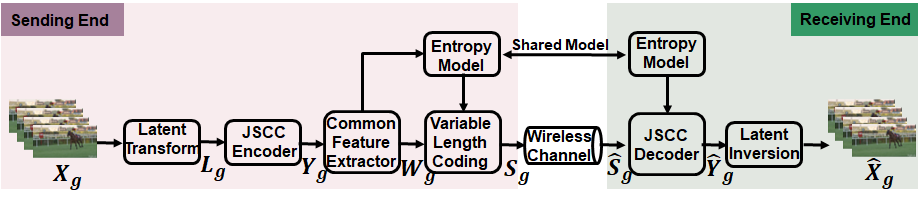}
	\caption{Diagram of VST Module}
	\label{fig:mdvsc}
\end{figure}

\label
UUser-provide videos with characters, remote background videos, and synthesized 3D videos need to be wirelessly transmitted between terminal devices (such as VR, 4K cameras, etc.) and edge server nodes. Limited wireless resources lead to long transmission delays, which has become a bottleneck for the current status of Metaverse immersive services. As shown in Fig. \ref{fig:mdvsc}, in order to reduce the delay of the VR video transmission module, we introduced semantic communication technology and deployed the semantic communication codec on edge nodes and end-side devices respectively.

 We adopt novel end-to-end semantic communication model proposed by Z.Bao et al.\cite{wang2003multiscale} to extract and restore semantic information. Take the real-time video $X^{i}$ transmitted from the end layer to the edge layer computing nodes as an example. 
In the semantic communication system, a video sequence $X$ to be transmitted is divided into Groups of Pictures (GOPs), with each GOP represented as $X_{g}$. As shown in Fig. \ref{fig:mdvsc}, The system first transforms $X$, into latent space to form latent representation $L_{g}$. This reduces data dimension and speeds up computation for large - scale spatial data. Then, the JSCC Encoder extracts semantic features from $L_{g}$, to output semantic feature maps $Y_{g}$. These feature maps are fed into the Common Feature Extractor (CFE), which extracts common and individual features in the GOP. lt generates onecommon feature map $W_{ge}$ and $N$ individual feature maps $W_{gi}$  for all frames. The combined feature maps are referred to as $W_{g}$. This process is formulated as:
\begin{equation}
	W_{gi} = Y_{g}-W_{gc} , W_{gc} = f_{c}(Y_{g})
\end{equation}
\begin{equation}
	W_{g}=(W_{gi},W_{gc})
\end{equation}
 where $f_{c}$(·) is the function for generating common features.
 
 An entropy model is applied to measure the information entropy of the symbol and restrict the data distribution of the CFE:
\begin{equation}
	em = f_{e}(W_{g}|d_{\alpha, \theta})
\end{equation}
 where $f_{e}$(·) is the probability mass function,$em$ denotes the likelihood, and $\theta$ is the weight and bias parameter in the deep entropy model.
 The final process at the transmitter is variable - length coding. Based on $W_{g}$, and its entropy, some elements are dropped to meet constraints like bandwidth cost. The output of variable - length coding is $S_{g}$. Symbols $S_{g}$ are distorted through the wireless channel. At the receiver, the JSCC decoder inverts the process of the JSCC encoder and outputs latent representation $\hat{Y}_{g}$. Finally, the latent inversion module transforms $\hat{Y}_{g}$, back to its original distribution space to generate recovered GOP $\hat{x}_{g}$, from which the output video $\hat{X}^{i}$ is obtained.

\subsection{Video Synthesis (VS) Module}\label{AA}
In order to realize the synthesis video of the virtual human image and the target background video, the video synthesized module, deployed in the cloud,  uses the efficient background matting network MODNet proposed by Ke et al.\cite{kingma2014adam} to extract the foreground from user-provide video. And then fuses the foreground human video with the background video $\hat{B}^{i}$.  

\begin{figure}[tbph]
	\centering
	\includegraphics[width=1\linewidth]{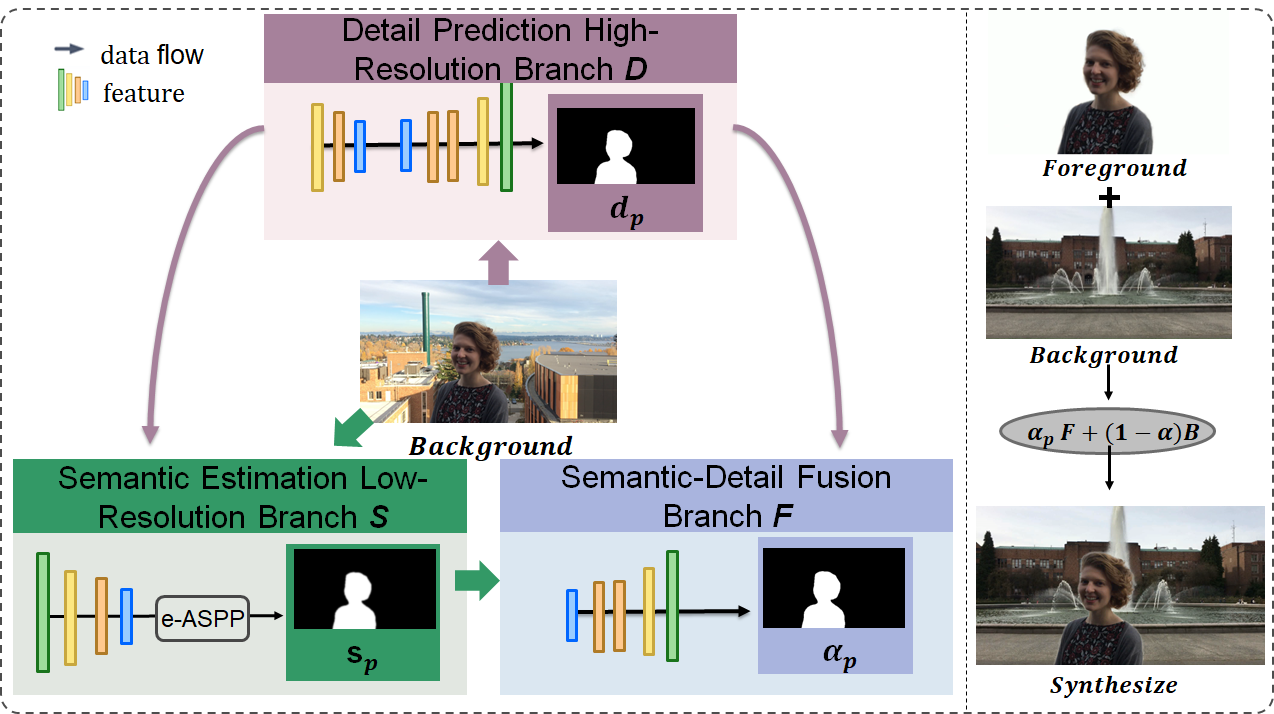}
	\caption{Diagram of VS Module}
	\label{fig:modnet-3}
\end{figure}
As shown in Fig. \ref{fig:modnet-3}, aiming at obtaining accurate semantic context, the network divides the trimap-free matting objective into three parts: semantic estimation, detail prediction, and semantic-detail fusion. And then, optimize them simultaneously via three branches.
Semantic Estimation is used to locate the position of the portrait, and we extract high-level semantics through an encoder which is the low-resolution branch $S$ of MODNet. To predict coarse semantic mask \( s_p \), we feed the high-level representation $S(I)$ into a convolutional layer activated by the sigmoid function. We supervise $s_{p}$ by a thumbnail of the ground truth matte \(\alpha_g \) . We use the $L_{2}$ loss as: 
\begin{equation}
	\zeta_s=\tfrac{1}{2}\begin{Vmatrix}s_{p}-G(\alpha _{g})
	\end{Vmatrix}_{2}
\end{equation}
High-resolution branch $D$ takes $I$, $S(I)$,and the low-level features from $S$ as inputs. The purpose of reusing the low-level features is to reduce the computational overhead of $D$. We denote the outputs of $D$ as $D(I; S(I))$.We calculate the boundary detail matte $d_{p}$ from $D(I; S(I))$ and learn it through the $L_{1}$ loss as:
\begin{equation}
	\zeta_d=m_{d}\begin{Vmatrix}d_{p}-G(\alpha _{g})
	\end{Vmatrix}_{1}
\end{equation}
where $m_{d}$ is a binary mask to let Ld focus on the portrait
boundaries. $m_{d}$ is generated through dilation and erosion on
$\alpha_{g}$. Its values are 1 if the pixels are inside the transition region, and 0 otherwise.

Branch $F$ combines the output of branch $D$ and branch $S$ to predict the graph, and the loss is as follows :
\begin{equation}
	\zeta_\alpha =\begin{Vmatrix}\alpha _{p}-\alpha _{g}
	\end{Vmatrix}_1-\zeta_{c}
\end{equation}
To sum up, we predicts portrait semantics $s_{p}$, boundary details $d_{p}$, and final alpha matte $\alpha_{p}$ through three interdependent branches, $S$, $D$, and $F$, which are constrained by explicit supervisions generated from the ground truth matte $\alpha_{g}$. 

After obtaining the character mask $\alpha$ through the designed network structure, we overlay user-provide video with the background video according to the mathematical model \(I^{i} =\alpha \hat X^{i}+(1-\alpha)\hat B^{i} \) to obtain the output synthesis video $I^{i}$.
 
\subsection{3D Virtual Scene Reconstruction (VSR) Module}

\begin{figure}[tbph]
	\centering
	\includegraphics[width=0.7\linewidth]{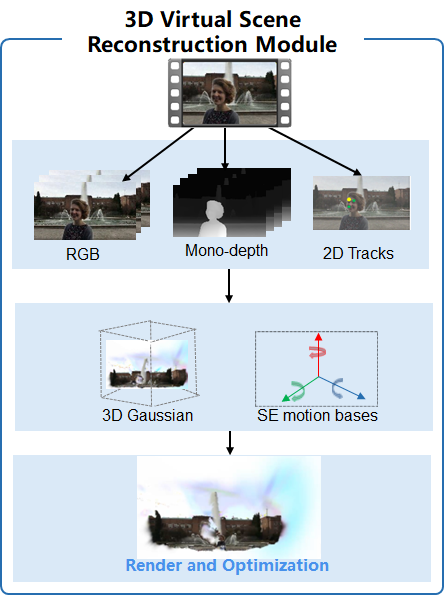}
	\caption{Diagram of VR Module}
	\label{fig:22}
\end{figure}
The Virtual Scene Reconstruction Module receives the synthesized video from the edge server via optical fiber and performs pre-processing in the cloud, which extracts depth information from the fusion video $I^{i}$, and renders the monocular 2D video into a 3D stereoscopic scene video by extracting the depth information and the person's movement trajectory\cite{wang2022meta}.
The RGB video can be shown as sequence $ \begin{Bmatrix}I_{t}\in \mathbb{R}^{H\times W\times 3}
\end{Bmatrix}$ with known camera poses $K_{t}\in \mathbb{R}^{3\times 3}$, along with monocular depth maps and 2D tracks computed from off-the-shelf models as input, we optimize a dynamic scene representation as a set of persistent 3D Gaussians that translate and rotate over time.
\begin{equation}
		\begin{aligned}
\mu^{'} _{0}(K,E)=\Pi(KE\mu _{0})\in \mathbb{R}^{2}, \\
\Sigma^{'}_{0}(K,E)=J_{KE}\Sigma_{0}J^{\top }_{KE}\in\mathbb{R}^{2\times 2}
	\end{aligned}
\end{equation}
\begin{equation}
\hat {\mathbf {I}}(\mathbf p)=\sum_{i\in H(p)}T_{i}\alpha _{i}c_{i}, \hat {\mathbf {D}}(\mathbf p)=\sum_{i\in H(p)}T_{i}\alpha _{i}d_{i}
\end{equation}
 To capture the low-dimensional nature of scene motion, we model the motion with a set of compact $\mathbb{S}\mathbb{E}(3)$ motion bases shared across all scene elements. Each 3D Gaussian’s motion is represented as a linear combination of these global $\mathbb{S}\mathbb{E}(3)$ motion bases, weighted by motion coefficients specific to each Gaussian. In particular, for a moving 3D Gaussian at time $t$, its pose parameters $(\mu _{t},R _{t}) $ are rigidly transformed from the canonical frame $t_{0}$ to $t$ via \( T_{0\to t} =\begin{bmatrix}R_{0\to t}T_{0\to t}
 \end{bmatrix}\in \mathbb{S}\mathbb{E}(3)  \)
 \begin{equation}
 \mu _{t}=R_{0\to t}+t_{0\to t}, (R_{t})=R_{0\to t}R_{0}
\end{equation}
where $\alpha_{i}= o_{i} \cdot  exp(-\frac{1}{2}(p-\mu _{0}^{'})), T_{i} = \prod_{j=1}^{i=1} (1-\alpha_{j})$, $H(p)$ is the set of Gaussians that intersect the pixel $p$ at query time $t$.
 The 2D correspondence location at time $t$ for a given pixel $p$,
 $ \hat{U}_{t\to t^{'}}(p)$, and the corresponding depth value at time $t^{'}$,$\hat{D}_{t\to t^{'}}(p)$can then be written as
 \begin{equation}
 \hat{D}_{t\to t^{'}}(p)=\Pi (K_{t^{'}}^{c}\textrm{}\hat{T}_{t\to t^{'}}(p)),\hat{D}_{t\to t^{'}}(p)=(^{c}\hat{X}_{t\to t^{'}}(p))_{[3]} 
 \end{equation}
 where $^{c}\hat{X}_{t\to t^{'}}(p)=E_{t^{'}}^{w}\hat{X}_{t\to t^{'}}(p)$.
  From above steps we soft decomposes the scene into multiple rigid moving groups to capture the low-dimensional motion structure of the scene, and performs dynamic new perspective synthesis and tracking rendering by optimizing the 4D Gaussian distribution function, ultimately achieving reconstruction of general dynamic scenes with clear, full-sequence long 3D motion from monocular synthesized video.

 \begin{figure*}
	\begin{center}
		\subfloat[]{
			\includegraphics[width=0.3\textwidth]{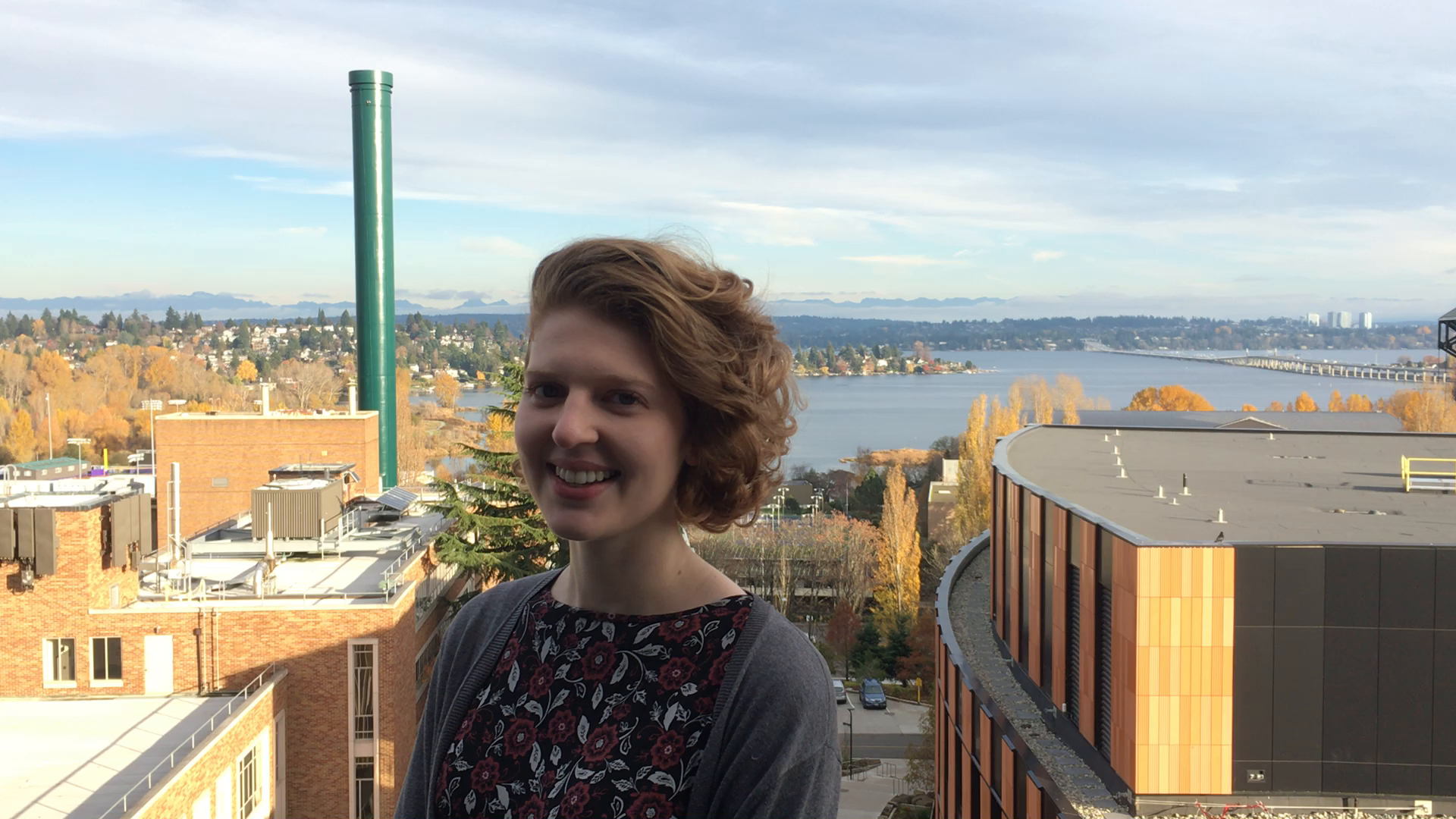}
		}
		\subfloat[]{
			\includegraphics[width=0.3\textwidth]{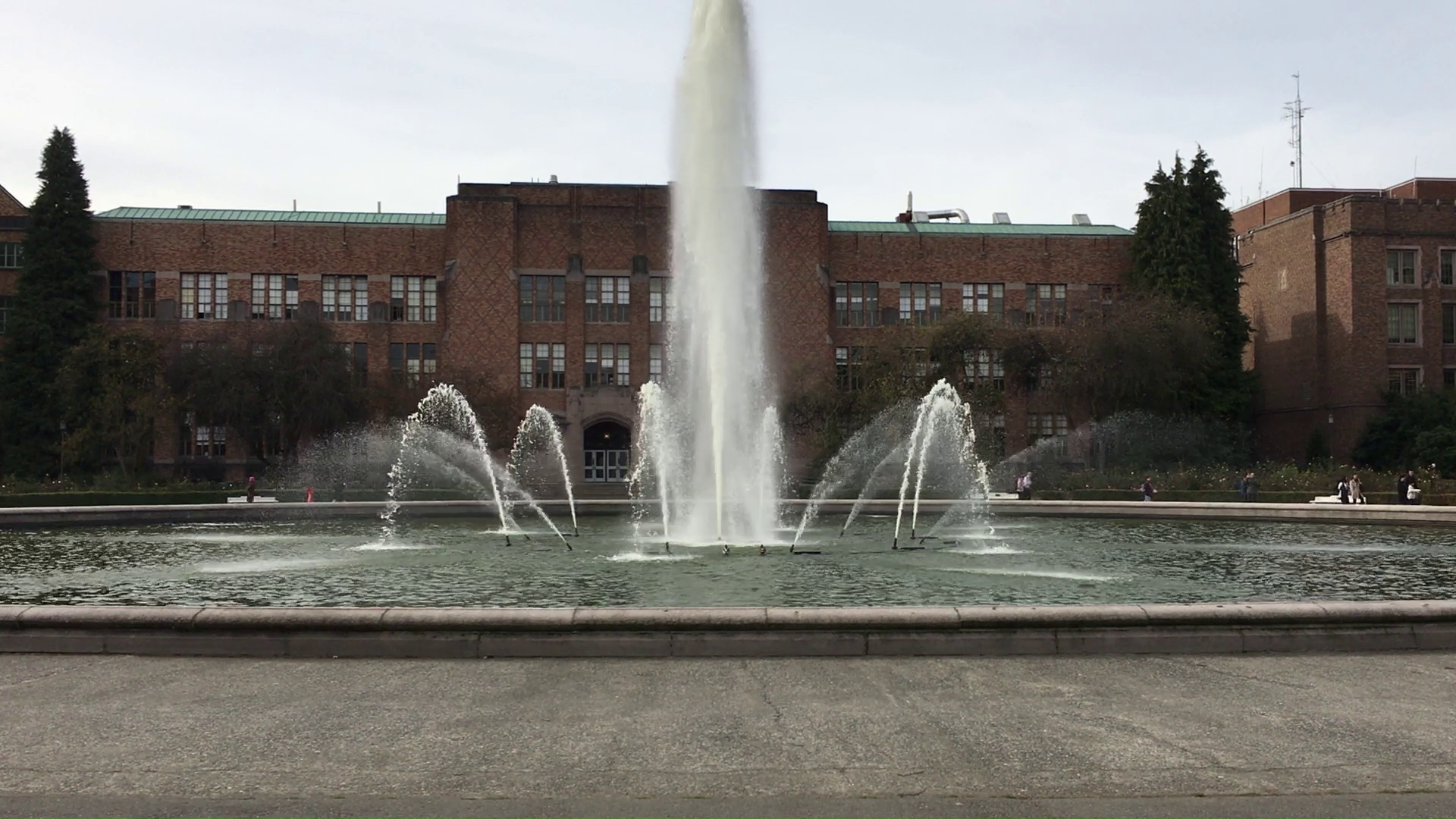}
		}
		\subfloat[]{
			\includegraphics[width=0.3\textwidth]{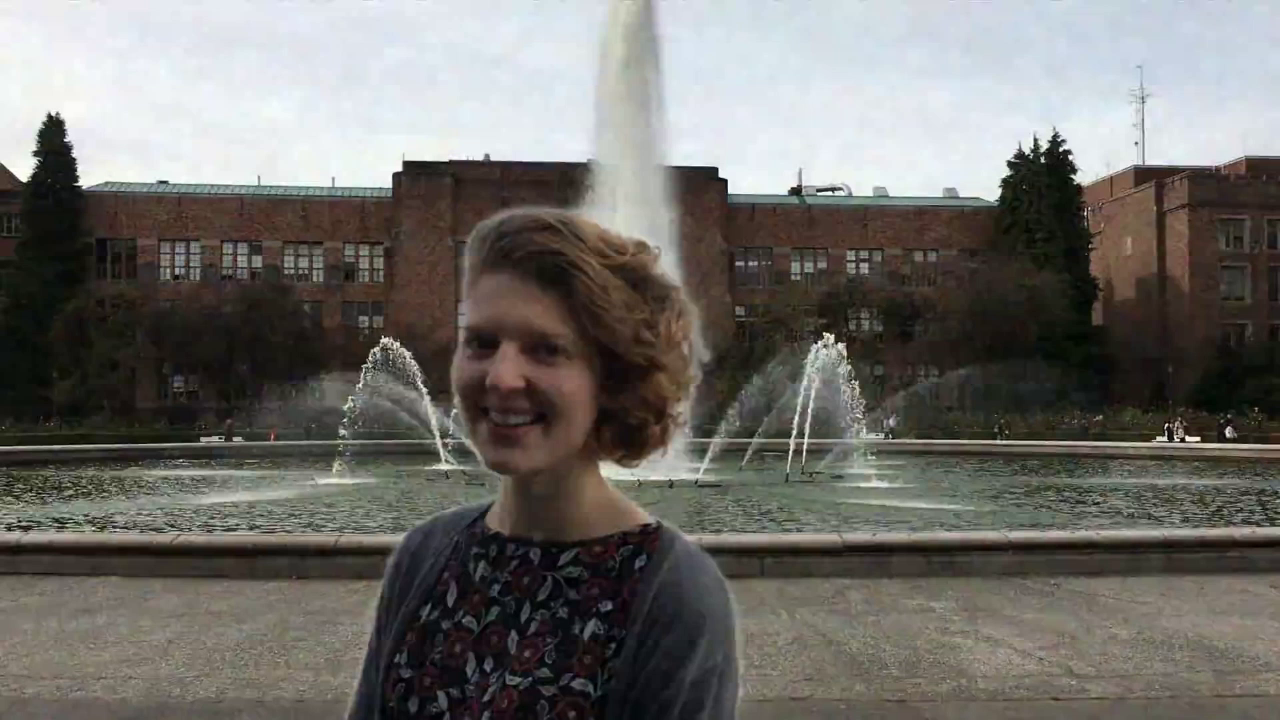}
		}
		
	\end{center}
	\begin{center}
		\subfloat[]{
			\includegraphics[width=0.3\textwidth]{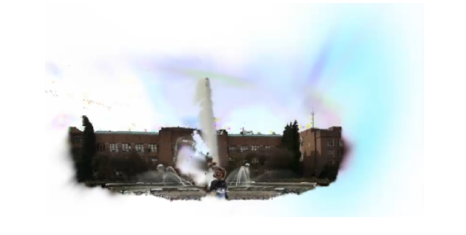}
		}
		\subfloat[]{
			\includegraphics[width=0.3\textwidth]{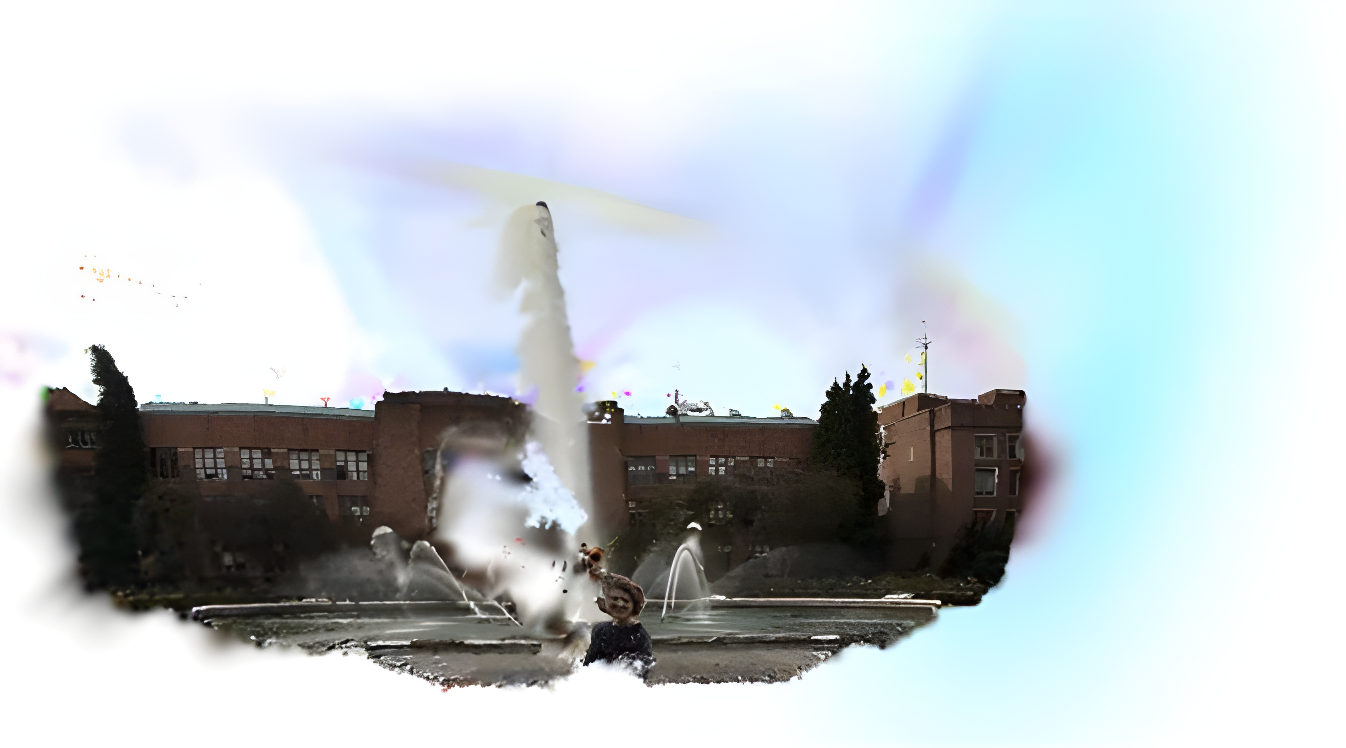}
		}
		\subfloat[]{
			\includegraphics[width=0.3\textwidth]{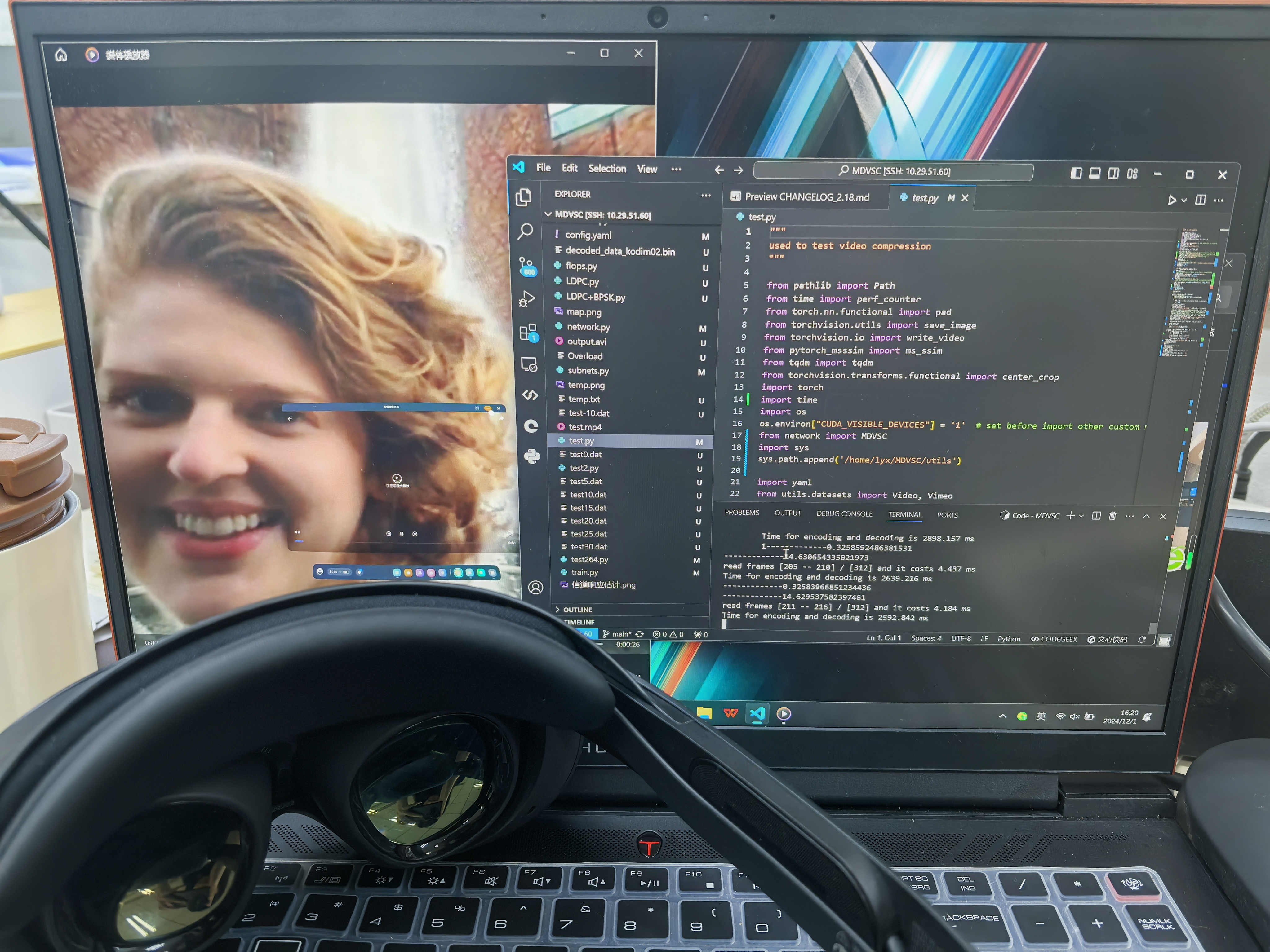}
		}
		\caption{SE-CEE-Meta Architecture Visualization.(a) User-provide video. (b) Background video. (c)Synthesis video. (d) (e) Multi-angle of 3D virtual reconstruction scene. (f)3D Video Displayed on VR Device.}
	\end{center}
		\label{fig:5s}
\end{figure*}
\section{EXPERIMENTS}
We conducted performance tests on the proposed framework to verify its functions and to demonstrate its advantages over existing metaverse services in terms of service latency on the wireless side as well as service quality under different channel conditions.
 \subsection{Simulation Settings}

1) Implementation Details

The built cloud-edge-end framework is implemented on the Python platform. The computing power settings of the cloud and edge are shown in the TABLE.\ref{tab:compute} . The computing power setting is based on the convention \cite{gu2022reliability}. The end-side device we choose meta quest pro. The size of the computing power consumed by 3D scence reconstruction module is about 60PFLOPS, among which the data preprocessing part in the cloud requires about 53 PFLOPS, the rendering part in the edge requires about 7 PFLOPS. Meanwhile, the VR video synthesized modules deployed on the cloud require about 50TFLOPS. So the computing power setting is relatively reasonable.

 \begin{table}[htbp]
	\caption{Requirements for computing power at the cloud, edge and side}
	\begin{center}
		\begin{tabular}{|c|c|c|c|}
			\hline
			&Provided Compute & Required Compute&\makecell[c]{Required \\Bandwidth}\\
			\hline
			Cloud&\makecell[c]{13-4000PFLOPS\\ Cloud center} &\makecell[c]{60PFLOPS\\(render)} &--  \\
			\hline
			Edge&\makecell[c]{ 10-100\newline TFLOPS\\(ZTE)}&\makecell[c]{100TFLOPS\\(video fusion)} &--\\	
			\hline		                   
			Side &\makecell[c]{ 1.2-1.5TFLOPS\\(meta quest pro)}&\makecell[c]{0.3-1TFLOPS\\(semantic extraction)} &\makecell[c]{200-800Mbps\\(8K video)} \\
			\hline
			
		\end{tabular}
		\label{tab:compute}
	\end{center}
\end{table}

In VST module, the channel dimension allocated for latent representations and JSCC (Joint Source-Channel Coding) codewords is set at 128. The VST module is compared with the classical video coding transmission schemes, including source coding H.264, channel coding 1/2LDPC [30], modulation BPSK for noise resistant and transmission efficiency respectively.

 In the VS module,we employed the PPM-100 dataset, which contains 100 finely annotated portrait images with various backgrounds to train network. The dataset was partitioned into training, validation, and testing sets, and the images were resized to 1920×1080 to facilitate processing.
 
In VSR module, we optimize our model using Adam Optimizer \cite{b23}. We perform 1000 iterations of optimization for the initial fitting and 500 epochs for joint optimization, respectively. The number of $\mathbb{S}\mathbb{E}(3)$ bases $B$ is set to 20 for all of our experiments\cite{b19}. Training on a sequence of 300 frames of 1920×1080 resolution takes about an hour to finish on NVIDIA GeForce RTX 4080 and NVIDIA RTX A5000. Our rendering FPS is around 40 fps.

2) Evaluation Metrics

The semantic communication system undergoes training with respect to the mean square error (MSE). During the testing phase, the evaluation is carried out based on the Peak Signal-to-Noise Ratio (PSNR) and  Multi-Scale Structural Similarity Index (MS-SSIM)\cite{b22} to assess the perceptual quality. The formulations for calculating these three metrics are as follows: 

\begin{equation}
	MSE(X,Y)=\frac{1}{mn}\sum_{m-1}^{i=0}\sum_{j=0}^{n-1}[X(i,j)-Y(i,j)]
\end{equation}
where $m$ and $n$ denote the number of pixels horizontally and vertically.
\begin{equation}
	\begin{aligned}
		PSNR(X,Y)=&10\cdot log_{10}\frac{1}{MSE}, \\
		MS-&SSIM(X,Y)= \\
		\left [ \zeta _{M}(X,Y) \right ]^{\alpha M}\cdot \prod_{j=1}^{M}&\left [C_{j}(X,Y)\cdot S_{j}(X,Y)  \right ]^{\alpha j}
	\end{aligned}
\end{equation}

In video synthesis module, we comprehensively evaluate the quality of the reconstructed scene from four aspects: 3D tracking, 2D tracking, image quality and perceptual quality. In terms of 3D tracking, EPE is an indicator that measures the accuracy of 3D point cloud tracking which calculates the Euclidean distance between the predicted 3D point position and the true position.
\begin{equation}
	EPE=\frac{1}{N}\sum_{i=1}^{N}\left\| p^{pred}_{i}-p^{gt}_{i}\right\|
\end{equation}
PCK measures the proportion of predicted points within the tolerance distance $\tau$. 
\begin{equation}
	PCK=\frac{1}{N}\sum_{i=1}^{N}\mathbb{I}(\left\|p^{pred}_{i}-p^{gt}_{i}\right\|<\tau )
\end{equation}
Jaccard Index evaluates the ratio of the intersection and union of the predicted box and the true box.
\begin{equation}
	AJ(\tau ) =\frac{1}{N}\sum_{i=1}^{N}\frac{\left|B^{pred}_{i}\cap B^{gt}_{i} \right|}{\left|B^{pred}_{i}\cup B^{gt}_{i} \right|}
\end{equation}
We use NV mPSNR and NV mSSIM to calculate the average of multiple frames of video based on PSNR and SSIM, which is suitable for neural networks.

 \subsection{Simulation Results and Analysis}
\begin{figure}[htbp]
	\centering
	\begin{minipage}{0.49\linewidth}
		\centering
		\includegraphics[width=0.9\linewidth]{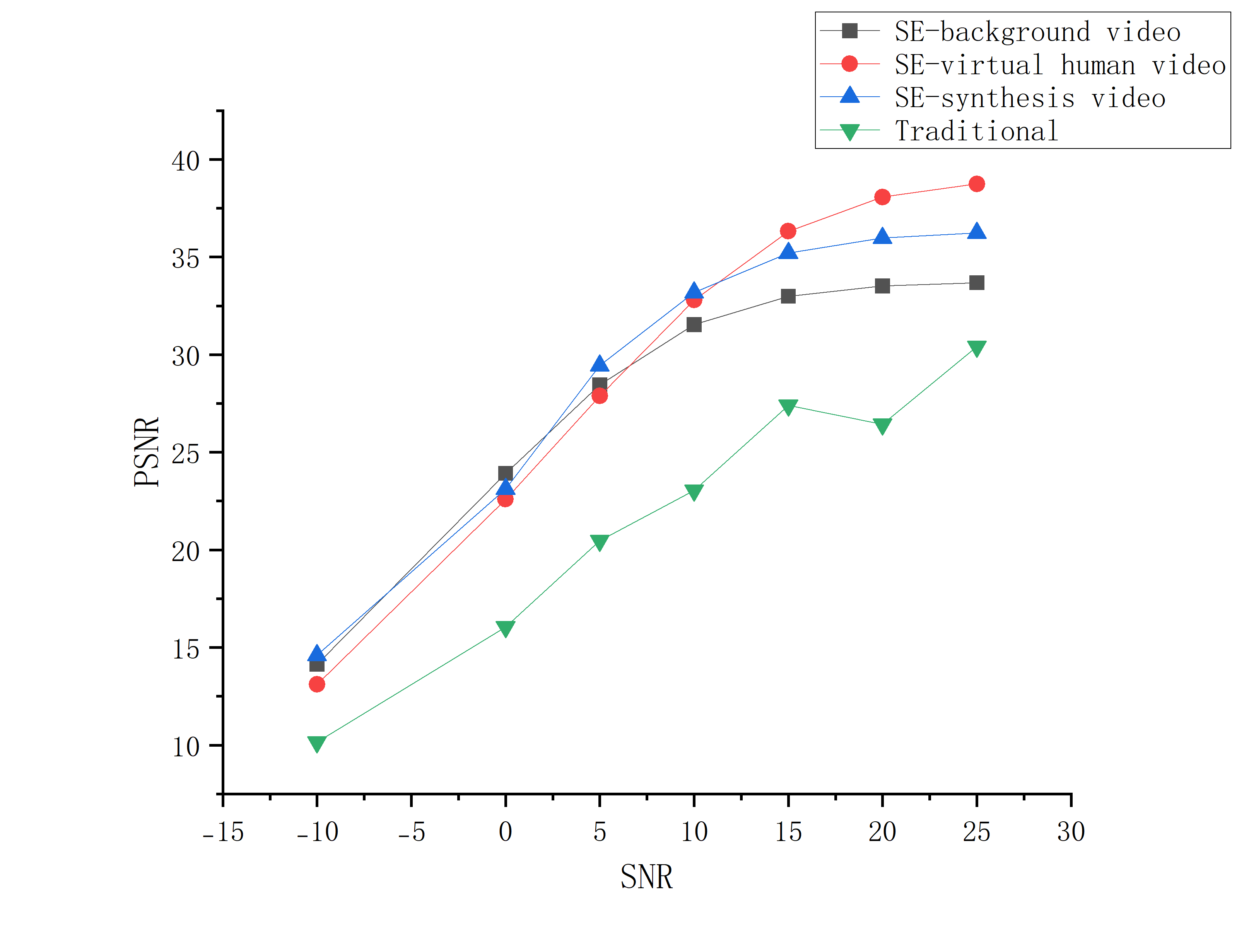}
		\caption{PSNR}
		\label{PSNR4}%文中引用该图片代号
	\end{minipage}
	%\qquad
	\begin{minipage}{0.49\linewidth}
		\centering
		\includegraphics[width=0.9\linewidth]{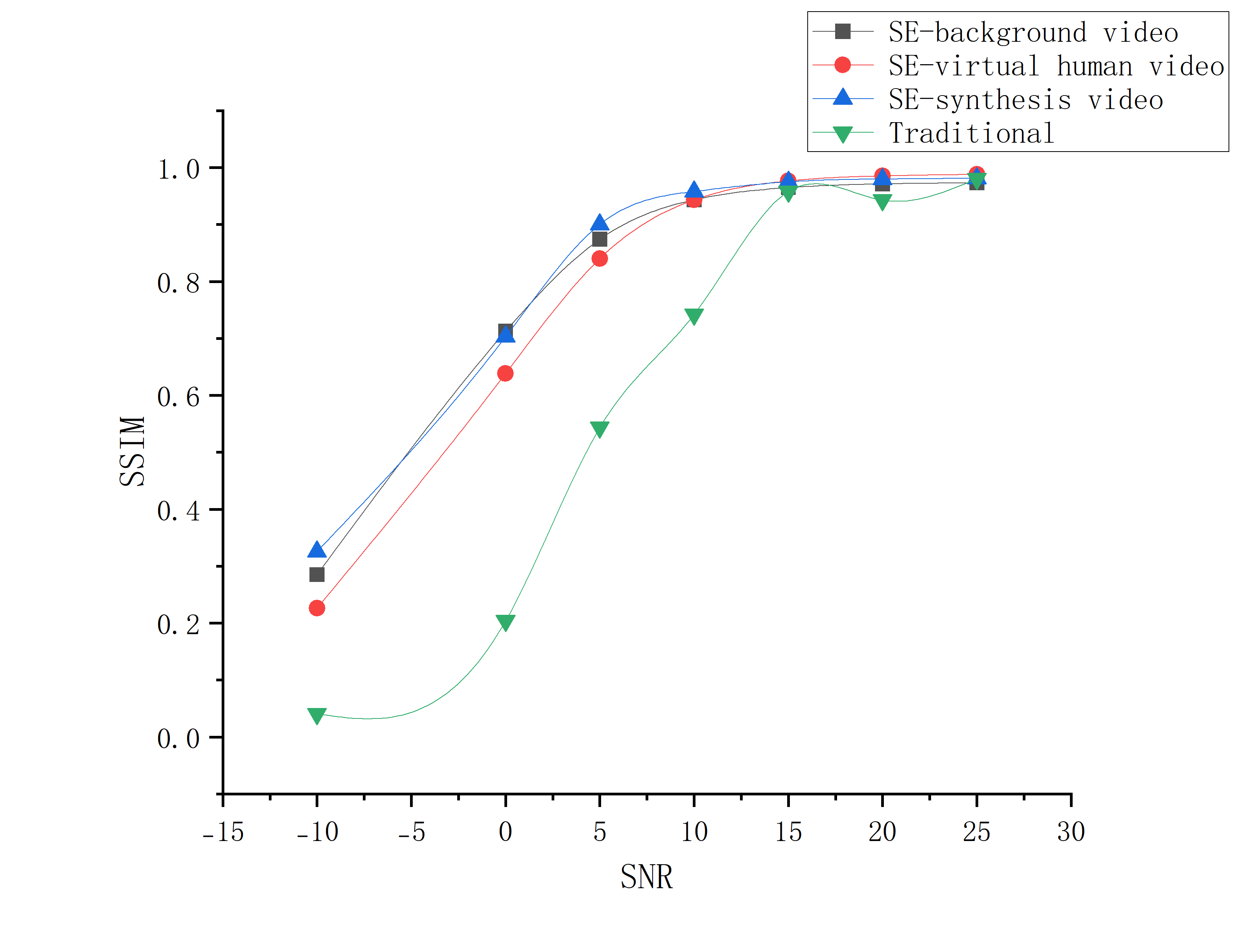}
		\caption{SSIM}
		\label{SSIM4}%文中引用该图片代号
	\end{minipage}
\end{figure}

1) The visualization effect of SE-CEE-Meta architecture is shown in the Fig. 5. We demonstrated the synthetic video received after semantic encoding and decoding on the VR device meta quest pro. The PSNR and MS-SSIM performance curves are shown in Fig.\ref{PSNR4} and Fig.\ref{SSIM4}, and the signal-to-noise ratio setting range is -10dB~25dB. The overall performance of the VST module is better than the anti-noise combination BPSK+1/2LDPC. Under poor channel quality (0dB), the image quality of the VST module can be improved by 43.99\% compared with the traditional communication method.
The transmission delay of semantic communication on the wireless side is significantly lower than that of traditional communication. 
 As shown in TABLE. \ref{tab3}, the delay of semantic communication is reduced by 5491 seconds compared with the average delay of traditional communication, which is about 96.03\%.
 
 Semantic communication optimizes the information transmission method and reduces unnecessary transmission data by using the redundancy of semantic levels. Thus, semantic communication performs better than traditional communication methods in terms of PSNR and SSIM, which means semantic communication can provide better communication quality even under poor channel quality.

\begin{table}[htbp]
	%\caption{Table Type Styles}
	\caption{Wireless Transmission Delay Comparison}
	\begin{center}
		\begin{tabular}{|c|c|c|c|c|}
			\hline
			&\makecell[c]{SE-virtual\\ human \\video}&\makecell[c]{SE-back-\\ground\\ video}&\makecell[c]{SE-synthesis\\ video}& \makecell[c]{TA-synthesis\\ video } \\
			\hline
			\makecell[c]{Wireless \\Transmission\\ Delay(s)}&463.48&455.343& 227.322&5718  \\
			\hline
			\makecell[c]{Video \\Size(MB)}&26.0&23.8& 11.5&11.5 \\
			\hline
		\end{tabular}
		\label{tab3}
	\end{center}
\end{table}

 2) As shown in TABLE. \ref{tab:shape}, in terms of 3D tracking, the Endpoint Error (EPE) error is small, and the Percentage of Correct Keypoints (PCK) can maintain a high accuracy when the threshold is 10cm and 5cm. There is jitter in 2D tracking and the video quality is reduced. The results show that the proposed module has certain advantages in 3D tracking, but there is some room for optimization in the average jitter and video quality of 2D tracking.
\begin{table}[htbp]
	\caption{3D Scene Reconstruction Performance Evaluation}
	\begin{center}
		\begin{tabular}{|c|c|c|}
			\hline
			Category& Indicator Name&Result Value   \\
			\hline
			3D Tracking&3D Tracking EPE&0.0337   \\
			\hline
			&3D Tracking PCK(10cm)&0.9933\\
			\hline
			&3D Tracking PCK(5cm)& 0.772  \\
			\hline
			2D Tracking&2D Tracking AJ&0.45   \\
			\hline
			Video quality&NV mPSNR&19.4357\\
			\hline
			&NV mSSIM& 0.5491  \\
			\hline
			&NV mLPIPS& 0.203  \\
			\hline
		\end{tabular}
		\label{tab:shape}
	\end{center}
\end{table}

\section{CONCLUSION}
This paper proposes a new Metaverse semantic communication framework (SE-CEE-Meta) to improve the effectiveness and efficiency of wireless Metaverse construction. By incorporating semantic communication, VR video synthesis, and virtual scene reconstruction into Metaverse construction, our proposed MetaSeco reduces the transmission delay by 96.03\% compared with the traditional wireless side communication framework.

\bibliographystyle{plain}
\bibliography{references.bib}

\end{document}